\documentclass[epj]{svjour}
\usepackage{graphics}
\usepackage{url}
\usepackage{amsmath}

\usepackage{xcolor}
\usepackage[normalem]{ulem}
\definecolor{dark green}{rgb}{0.0, 0.5, 0.0}
\definecolor{dark red}{rgb}{0.5, 0.0, 0.0}
  
\begin{document}
\title{Verification of He-3 proportional counters fast neutron sensitivity through a comparison with He-4 detectors}
\subtitle{He-3 and He-4 proportional counters fast neutron sensitivity and evaluation of the cosmic neutron fluxes at ESS}
\author{Francesco Piscitelli\inst{1}\thanks{Corresponding author: francesco.piscitelli@ess.eu} 
\and Giacomo Mauri\inst{1,2} \and Alessio Laloni\inst{1} \and Richard Hall-Wilton\inst{1,3}
}                     

%
\institute{European Spallation Source ERIC, P.O. Box 176, SE-221 00 Lund, Sweden. \and Science and Technology Facilities Council, ISIS department, Rutherford Appleton Labs, Didcot OX11 0QX, UK. \and Dipartimento di Fisica ``G. Occhialini'', University of Milano-Bicocca, Italy.}
\date{Received: date / Revised version: date}
%
\abstract{In the field of neutron scattering science, a large variety of instruments require detectors for thermal and cold neutrons. Helium-3 has been one of the main actors in thermal and cold neutron detection for many years. Nowadays neutron facilities around the world are pushing their technologies to increase the available flux delivered at the instruments, this enables a completely new science landscape. Complementary with the increasing available flux, a better signal-to-background ratio (S/B) enables to perform new types of measurements. To this aim, this manuscript re-examines the background sensitivity of today's ``gold standard" neutron detection. Fast neutrons and gamma-rays are the main background species in neutron scattering experiments. The efficiency (sensitivity) of detecting fast neutrons, cosmic rays and gamma-rays, for an Helium-3-based detector is studied here through the comparison with Helium-4 counters. The comparison to Helium-4 allows to separate the thermal (and cold) neutron to the fast neutron contributions in Helium-3-based counters, which are otherwise entangled; verifying previous results from an indirect method. A relatively high sensitivity is found. Moreover, an estimate for the cosmic neutron fluence, also a source of background, at ground level at ESS is presented in this manuscript.
} 

\titlerunning{He-3 and He-4 proportional counters fast neutron sensitivity}
\authorrunning{F. Piscitelli et al.}

\maketitle
\section{Introduction}
\label{intro}

Nowadays neutron scattering facilities are increasing their available flux delivered at their instruments which, not only allows more complex scientific investigations, but also enables new science. The new generation of high-intensity neutron sources, such as the Japan Spallation Source (J-PARC ~\cite{JPARC}), the Spallation Neutron Source (SNS~\cite{SNS}) in the US, and the European Spallation Source (ESS~\cite{ESS,ESS_TDR,ESS-design,ESS2011}) presently under construction in Lund (Sweden), are able to provide, with respect to the previous generation sources, at least one order of magnitude higher flux at the instruments~\cite{ESS2011,Mezei2007,VETTIER_ESS,MIO_ESSInstrSuite}. Along with the gain of available and useful neutron flux, an increase of background is expected; i.e. a better signal-to-background ratio (S/B) is needed to have highly performing instruments. A scientific result depends on the S/B as a whole, a higher intensity of the source does not implicate a better result and the improvement of the background is also essential. For small signals the ability to extract a signal scales with S/B$^2$ which makes the background rejection even more crucial. 
\\ The background at an instrument can be partially reduced by shielding, anyhow heavy shielding components do not always represent a viable solution due to the limited space available and its high cost. Moreover, a background radiation is delivered through the same delivery system used to guide thermal, useful, neutrons to the instrument. Hence, in order to achieve a high S/B rejection, it is imperative for a detector for neutron scattering applications to require a high detection efficiency (at thermal and cold energies, $\approx$25~meV) and a very low efficiency (sensitivity) to other types of radiation. A source of background in a neutron scattering facility is mainly due to gamma-rays, epi-thermal, fast neutrons, cosmic rays and electronic noise~\cite{BG_cherkashyna2014}. When any of these radiations are detected, they result in spurious events recorded as true events. It is possible to identify background events only if a pulse analysis is performed, that usually is limiting the counting rate at which the detector can be operated. 

The sensitivity of a thermal neutron detector to fast neutrons (or gamma-rays) is defined as the probability for a fast neutron (or a gamma-ray) to generate a false count in a neutron measurement. Analytically the sensitivity is defined as the efficiency. The sensitivity (or efficiency) is the probability for an incident radiation on a detector element (a whole counter in the present work) to result into an event. Practically, if a portable source is used the sensitivity is the number of events, that exceed a set energy threshold in the Pulse Height Spectrum (PHS), normalised to the activity, intensity of the source and the solid angle subtended to the detector element. 
\\ The term sensitivity is used to differentiate from efficiency. Whereas the latter represents a valuable feature of a device, that is desirable to be as high as possible; the sensitivity is the efficiency of detecting an unwanted radiation: i.e. detecting a fast neutron (or a gamma-ray) that is misinterpreted as a thermal or cold neutron. Together, the efficiency and the sensitivity, define the best achievable S/B for a detector. 

Although Helium-3 has been the main actor in neutron detection for decades, many alternative technologies~\cite{HE3S_karl,HE3S_hurd,COLL_icnd} have been proposed and developed during the last ten years to face the challenge of the Helium-3 shortage and performance~\cite{HE3S_hurd,HE3S_cho,HE3S_gao,HE3S_kramer,HE3S_shea}. The Helium-3 performance in terms of counting rate capability is strictly connected to the way the detector is built. On the other hand, the spatial resolution and the sensitivity to background is a peculiar feature of the Helium-3 itself. Helium-3 still represents a valid neutron detection medium for scattering science for some applications. A valid alternative to Helium-3 is Boron-10~\cite{B4C_carina,B4C_carina3,B4C_Schmidt} used as a solid converter layer in detectors. The Multi-Grid~\cite{MG_2017,MG_IN6tests,MG_joni,MG_patent} for large area applications in neutron spectroscopy, the Multi-Blade~\cite{MIO_HERE,MIO_MB16CRISP_jinst,MIO_MB2014,MIO_MB2017,MIO_MBproc,MIO_MyThesis,MIO_ScientificMBcrisp} for neutron reflectometry, the Jalousie detector~\cite{DET_jalousie,DET_Jalousie3} for neutron diffraction, BandGEM~\cite{Bgem,MPGD_GEMcroci}, the Boron-coated straws~\cite{STRAW_athanasiades,STRAW_lacy2,STRAW_lacy2002,STRAW_lacy2006,STRAW_lacy2011,STRAW_lacy2013} for SANS instruments, and CASCADE~\cite{DET_kohli,MPGD_KleinCASCADE} are some of the newly developed Boron-10-based detectors. Moreover, other converter materials have been also investigated as a replacement for Helium-3, as an example Gadolinium has been coupled to a Gas Electron Multiplier device (GEM~\cite{DET_GEM1}) in the Gd-GEM detector~\cite{DET_doro1,gdgem} or coupled to solid state detectors~\cite{SCHULTE_sigd,Mireshghi_sigd,PETRILLO-solidstate}.

It has been shown that Boron-10-based proportional counters or Multi-Wire-Proportional-Chambers (MWPC) can achieve a gamma-ray sensitivity as low as Helium-3-based detectors below $10^{-6}$~\cite{MG_gamma,MIO_MB2017}. 

Boron-10-based detector fast neutron sensitivity has been measured and simulated and it is of the order of $10^{-5}$~\cite{MIO_fastn}. Indirect measurements~\cite{MIO_fastnhe3giac} and simulations have indicated that Helium-3 has a fast neutron sensitivity of the order of $10^{-3}$. 

In this manuscript the sensitivity to fast neutrons and gamma-rays of a Helium-3 counter is verified by comparing with an Helium-4 counter in order to disentangle the sensitivity to fast neutrons from the one to thermal neutrons. The theoretical considerations, given in the next section, will clarify why measuring the Helium-4 sensitivity to fast neutrons represents a valid approximation to extrapolate that of Helium-3.
\\ Note that epi-thermal neutrons also contribute to the background, but this contribution is not discussed in this manuscript. The fast neutron energy range is set by the available neutron sources at our laboratory and it ranges between 1 and 10\,MeV. Fast neutron background energy range in spallation sources is wider than this and can extend up to 100\,MeV. 
\\ Moreover, the measurement with cosmic rays, also a possible source of background, is shown here to verify the actual thermal and fast cosmic neutron fluxes present at the ground level at ESS.

\section{Theoretical considerations}
\label{theo}
The sensitivity of gaseous detectors to fast neutrons is mainly given by the gaseous media and very little contribution is observed from the solid materials of the detector~\cite{MIO_fastn}. In the case of Boron-10-based detectors, the neutron converting material can be easily removed from the detector to evaluate experimentally the detection of fast neutrons. On the other hand, in Helium-3 detectors the detection of thermal and cold neutrons cannot be disentangled from the detection of fast or epi-thermal neutrons. Simulations and indirect measurements have been performed on this matter~\cite{MIO_fastnhe3giac}. 
\\ Helium-3 is used as a means in fast neutron spectroscopy~\cite{DET_knoll,He3spectr,He3spectr2}, whereas Helium-4 is employed for fast neutron detection~\cite{DET_knoll,FastN1,FastN2,FastN3,FastN4,arktis,STF_Ram}. Helium-4 is not sensitive to thermal neutrons through a conversion process, hence it gives an insight of the fast neutron sensitivity of Helium-3 at fast neutron energies. With Helium-4, the fast neutron sensitivity can be disentangled from the sensitivity to thermal and cold neutrons of Helium-3. The possible interactions that a neutron can undergo when interacting with Helium-3 and Helium-4 nuclei are listed here:

\begin{table}[htbp]
\centering
\begin{tabular}{ll}
- $\mathrm{^3He}$  & \\
\qquad $\bullet$ $\mathrm{^3He}$ recoil (elastic scattering): & $\mathrm{^3He} + n \rightarrow \mathrm{^3He}' + n'$ \quad $\left(E_{R\,max} =0.75\,E_{i}\right)$\\
\qquad $\bullet$ $\left( n,p \right)$: & $\mathrm{^3He} + n \rightarrow \mathrm{^1H} + \mathrm{^3H} + 0.764\,MeV$ \\
\qquad $\bullet$ $\left( n,d \right)$: & $\mathrm{^3He} + n \rightarrow \mathrm{^2H}  + \mathrm{^2H} - 3.27\,MeV$ \\
\qquad $\bullet$ $\left( n,\gamma \right)$ $\gamma$-rays &  \\
\noalign{\smallskip}\hline\noalign{\smallskip}
- $\mathrm{^4He}$ & \\
\qquad $\bullet$ $\mathrm{^4He}$ recoil (elastic scattering): & $\mathrm{^4He} + n \rightarrow \mathrm{^4He}' + n'$ \quad $\left(E_{R\,max} =0.64\,E_{i}\right)$\\
\end{tabular}
\end{table}

\begin{figure}[htbp]
\centering
\begin{tabular}{ccc}
\resizebox{0.31\textwidth}{!}{\includegraphics{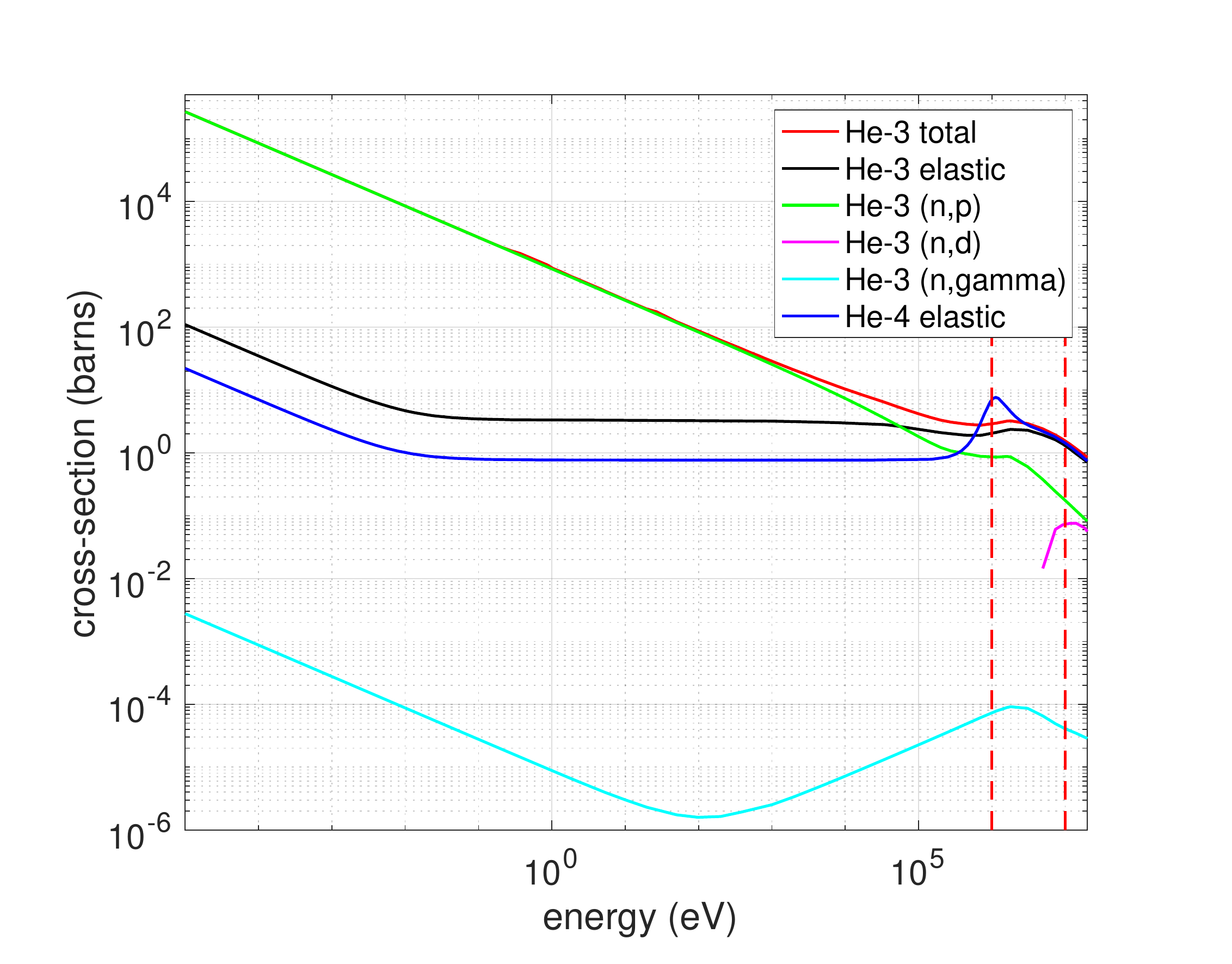}}&
\resizebox{0.31\textwidth}{!}{\includegraphics{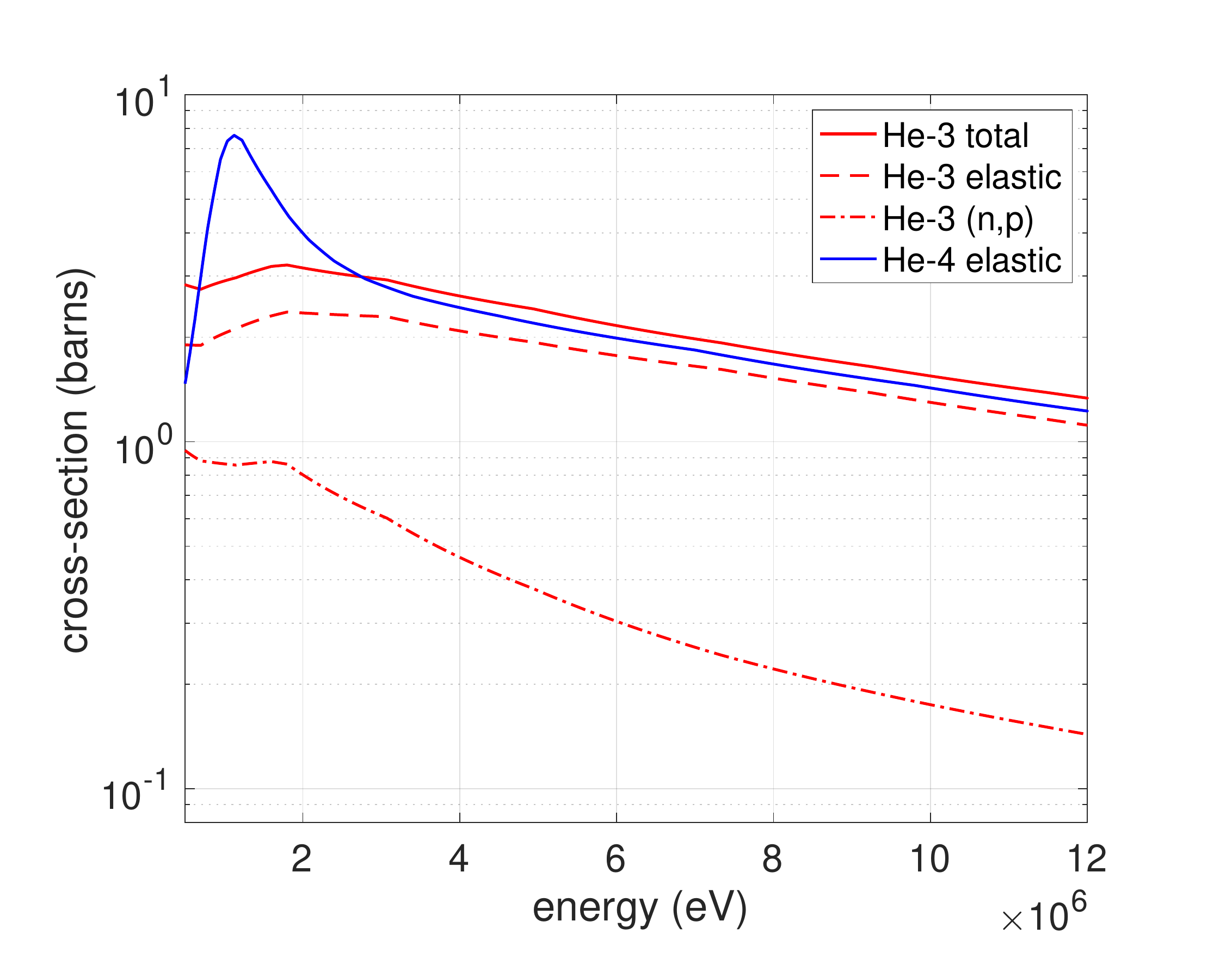}}&
\resizebox{0.31\textwidth}{!}{\includegraphics{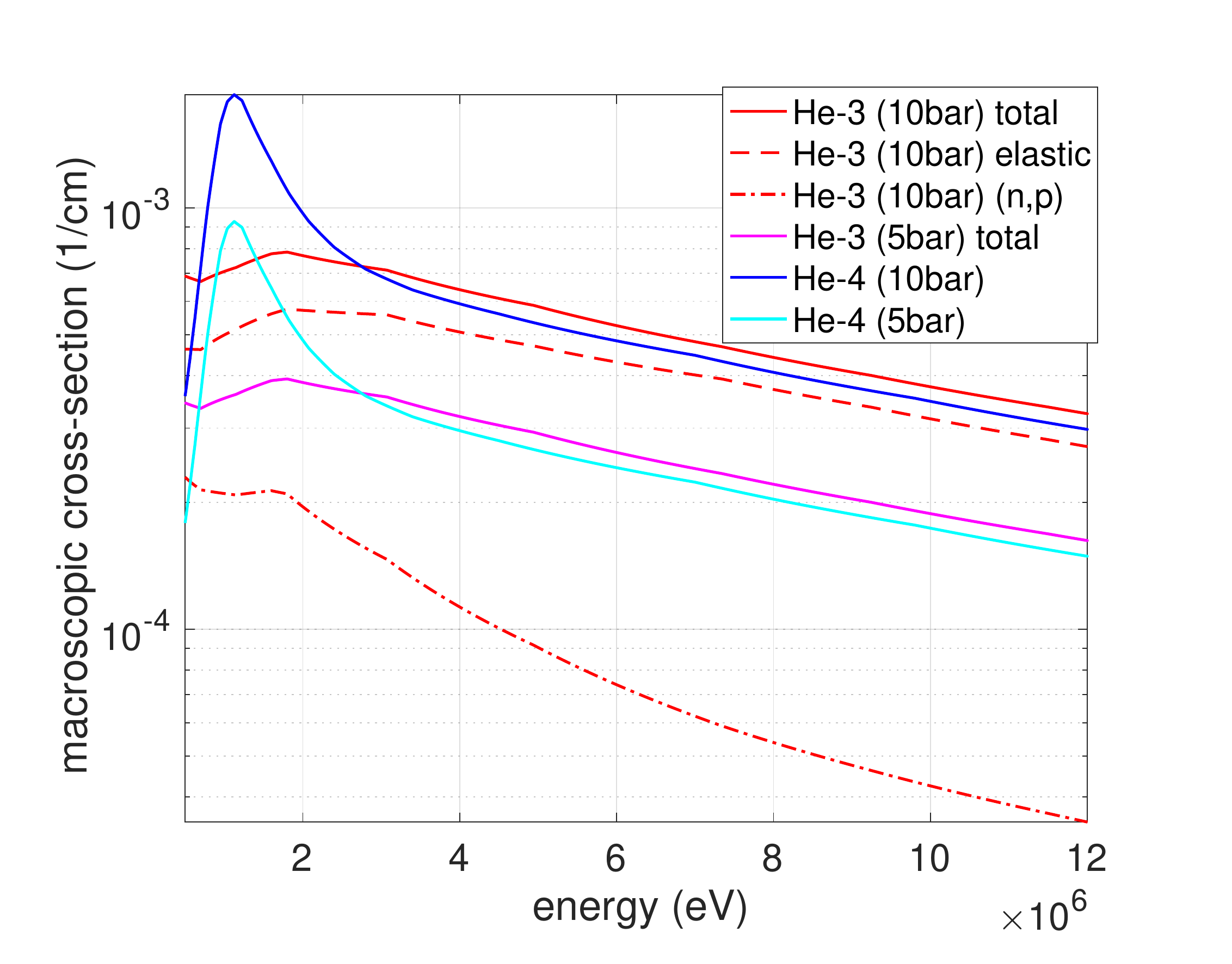}}\\
(a) & (b) & (c)
\end{tabular}
\caption{\footnotesize a) Cross-section for various processes of a neutron interacting with either Helium-3 or Helium-4 nuclei. b) Same as (a) in the neutron energy range 0.5-12~MeV. c) Macroscopic cross-sections for 5 and 10~bar of Helium-3 or Helium-4 in the energy range 0.5-12~MeV. From~\cite{NIST}.}
\label{figsigma} 
\end{figure}

\newpage
Helium-4, opposite to Helium-3, can interact with neutrons only via elastic scattering (recoil) in the whole energy range that spans from cold up to fast neutron energies. On the other hand, a neutron can be also captured (absorption interaction) or excite the target nucleus (inelastic interaction) when interacting with Helium-3. Figure~\ref{figsigma} shows the cross-sections of those processes, in the wide energy range from $\mu$eV up to MeV (a) and in the energy range 1-10~MeV (b) of interest for this manuscript. Moreover, the the macroscopic cross-section is shown distinguishing between two gas pressure for each type of detector (c).

In the case of neutrons interacting with Helium-3, the $\left( n,\gamma \right)$ process can be neglected being its cross-section of several orders of magnitude smaller than any other possible reaction. The $\left( n,d \right)$ reaction occurs at energies above $\approx$10~MeV. It can be neglected as well, since the source used for this investigation emits neutrons in the 1-10~MeV energy range.
Moreover, its cross-section is two orders of magnitude smaller than the cross-section of recoil (elastic interaction) and (n,p) (absorption) processes. Thus only these two latter mechanisms are considered for the calculations regarding Helium-3. 
\\ In the 1-10~MeV range the combination of absorption and recoil in Helium-3 sums up to a comparable cross-section to the sole recoil cross-section of Helium-4. The average total cross-section in the considered energy range is 2.54~barns and 2.34~barns for Helium-4 and Helium-3 respectively; i.e. the probability of interaction with Helium-3 and Helium-4 is similar. 
\\ For a given gas pressure, Helium-3 and Helium-4 have the same number density ($2.4\cdot10^{19}\,\mathrm{cm^{-3}}$ at 1~bar) and it scales linearly with their partial pressure. On the other hand, the mass density of the two gases differs and this impacts the way a Helium ion releases its energy into the gas volume. The mass density is $1.2\cdot10^{-4} \, \mathrm{g/cm^3}$ and $1.6\cdot10^{-4} \, \mathrm{g/cm^3}$ at 1~bar for Helium-3 and Helium-4 respectively. 
\\ Figure~\ref{figsigma} (c) shows the macroscopic cross-section ($\Sigma$) for the two gases at the pressures of 5 and 10~bar. The macroscopic cross-section represents the probability, per unit length, of interaction of a neutron with the nucleus under consideration; i.e. for small values of $\Sigma\cdot x$ the interaction probability can be approximated as:

\begin{equation}\label{eq2}
P(x)= 1-e^{-\Sigma\cdot x}\approx \Sigma \cdot x
\end{equation}

Hence the plot in figure~\ref{figsigma} (c) represents also the probability per unit length ($P(x)/x$) of having an interaction with any of the two nuclei for a given pressure, 5 and 10~bar in this specific case. At fixed partial pressure (e.g. at 10~bar) the interaction probability (given by the total cross-section) for Helium-3 and Helium-4 can be considered similar and of the order of $10^{-4}-10^{-3}/\mathrm{cm}$. For instance, for a 2.54~cm diameter tube filled with 10\,bar Helium, the probability of interaction is simply 2.54 times higher. 

\begin{figure}[htbp]
\centering
\resizebox{0.45\textwidth}{!}{\includegraphics{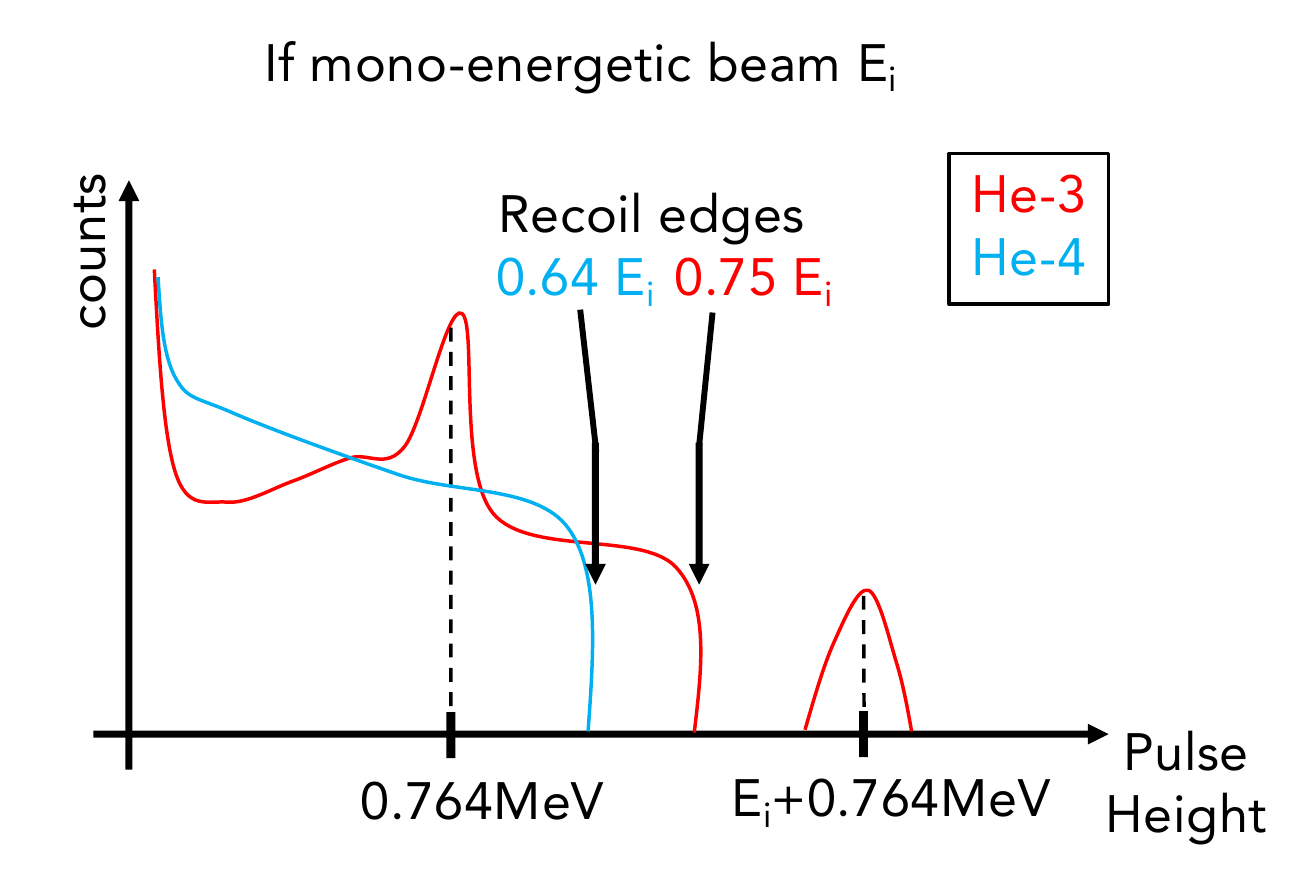}}
\caption{\footnotesize Theoretical sketch of the PHS obtained with a counter (neglecting wall-effects) filled with Helium-3 or Helium-4 when exposed to a mono-energetic $E_i$ fast neutron beam.}
\label{figtphstheo}
\end{figure}

In the case of an elastic interaction, either for Helium-3 or Helium-4, the maximum energy that the recoiled nucleus can receive in the interaction is given by equation~\ref{eq1}.

\begin{equation}\label{eq1}
E_{R\,max} = \frac{4A}{(1+A)^2}\,E_i
\end{equation}

where A is the mass number of the isotope. 
\\ Helium-3 detectors are used in fast neutron spectroscopy measurements~\cite{DET_knoll,He3spectr,He3spectr2}, because in the case of a mono-energetic fast neutron beam, the incoming neutrons energy can be deduced. Let us consider an ideal detector case, where the wall-effect can be neglected, and a beam of mono-energetic $E_i$ incoming neutrons. In the case of Helium-4 the cross-section is totally elastic and only recoil occurs. From equation~\ref{eq1}, the maximum energy transferred is $E_{R\,max} = 0.64\cdot E_i$. This gives a continuous distribution in the Pulse Height Spectrum (PHS) which spans from zero up to $E_{R\,max}$. Figure~\ref{figtphstheo} shows a sketch of a possible PHS obtained with a mono-energetic neutron beam for both isotopes. In the case of Helium-3, the absorption cross-section $(n,p)$ gives rise to a peak in the PHS at the energy $E_i+Q$, where $Q$ is the Q-value for the neutron capture reaction which is $Q=0.764\,\mathrm{MeV}$. The elastic cross-section contributes to a recoil distribution in the PHS which extends up to $E_{R\,max} = 0.75\cdot E_i$. 
\\ If also thermal and epi-thermal neutrons are also present, they both give rise to the characteristic peak at energy Q in the PHS due to the Helium-3 capture reaction. Thermal and epi-thermal neutron cannot be distinguished in the PHS. At energies below $\approx 10^4\,\mathrm{eV}$ the absorption cross-section exceeds the elastic cross-sections and the absorption process becomes much more probable than recoil. Therefore, the peak at energy Q becomes predominant at these energies~\cite{DET_knoll}. 

If a neutron source is used, the incoming energy distribution is far from mono-energetic and along with fast also thermal and epi-thermal neutrons will be present due to the moderation processes happening in the surroundings of the detector. The PHS for Helium-3 in this case will be the superposition of three components. The characteristic peak at energy Q due to thermal and epi-thermal neutrons, the summation of many recoil events each with maximum energy $E_{R\,max}$ and the overlap of many capture events each with energy $E_i+Q$. Thus no peak at $E_i+Q$ can be identified any more. On the other hand, for Helium-4 the PHS will be a simple continuum obtained by the summation of the several events given by the recoiled nuclei each with maximum energy $E_{R\,max}$.
\\ Moreover, in a real detector case, a detector has a finite pressure and volume which introduces the wall-effect in the PHS. 
\\ Once a nucleus recoils, it loses energy according to the Bethe-Block formula~\cite{DET_knoll}. The stopping power of a $\mathrm{^3He^{++}}$ ion in Helium-3 or $\mathrm{^4He^{++}}$ ion in Helium-4 for a given pressure and incoming energy can be calculated with SRIM~\cite{MISC_SRIM1998,MISC_SRIM2010}. The stopping power allows to extrapolate the range of those particle in the gas and the amount of energy that can be deposited in a given volume. Table~\ref{tab1} shows the ranges for several energies for both Helium-3 and Helium-4 ions for 1~bar pressure. These values scale linearly inverse with the pressure. 

\begin{table}[htbp]
\centering
\caption{\footnotesize Ranges of recoil nuclei for 100\,keV energy threshold and 1 bar. Relative uncertainties never exceed 3\%.}
\label{tab1}
\begin{tabular}{ccc}
\hline\noalign{\smallskip}
        energy (MeV)   & range (cm)  & range (cm) \\ 
                              & $\mathrm{^3He^{++}}$ in $\mathrm{^3He}$  & $\mathrm{^4He^{++}}$ in $\mathrm{^4He}$  \\ 
        \noalign{\smallskip}\hline\noalign{\smallskip}
        1 & 2.14 & 2.39 \\
        2 & 4.61 & 5.03 \\
        3 & 8.09  & 8.55 \\ 
        4 & 12.52 & 12.98  \\
        5 & 17.89 & 18.29 \\
        6 & 24.16 & 24.47 \\ 
        7 & 31.27 & 31.49 \\
        8 & 39.25 & 39.34\\
        9 & 48.06  & 47.96  \\ 
      10 & 57.60 & 57.29 \\ 
       \noalign{\smallskip}\hline
\end{tabular}
\end{table}

\begin{figure}[htbp]
\centering
\begin{tabular}{ccc}
\resizebox{0.31\textwidth}{!}{\includegraphics{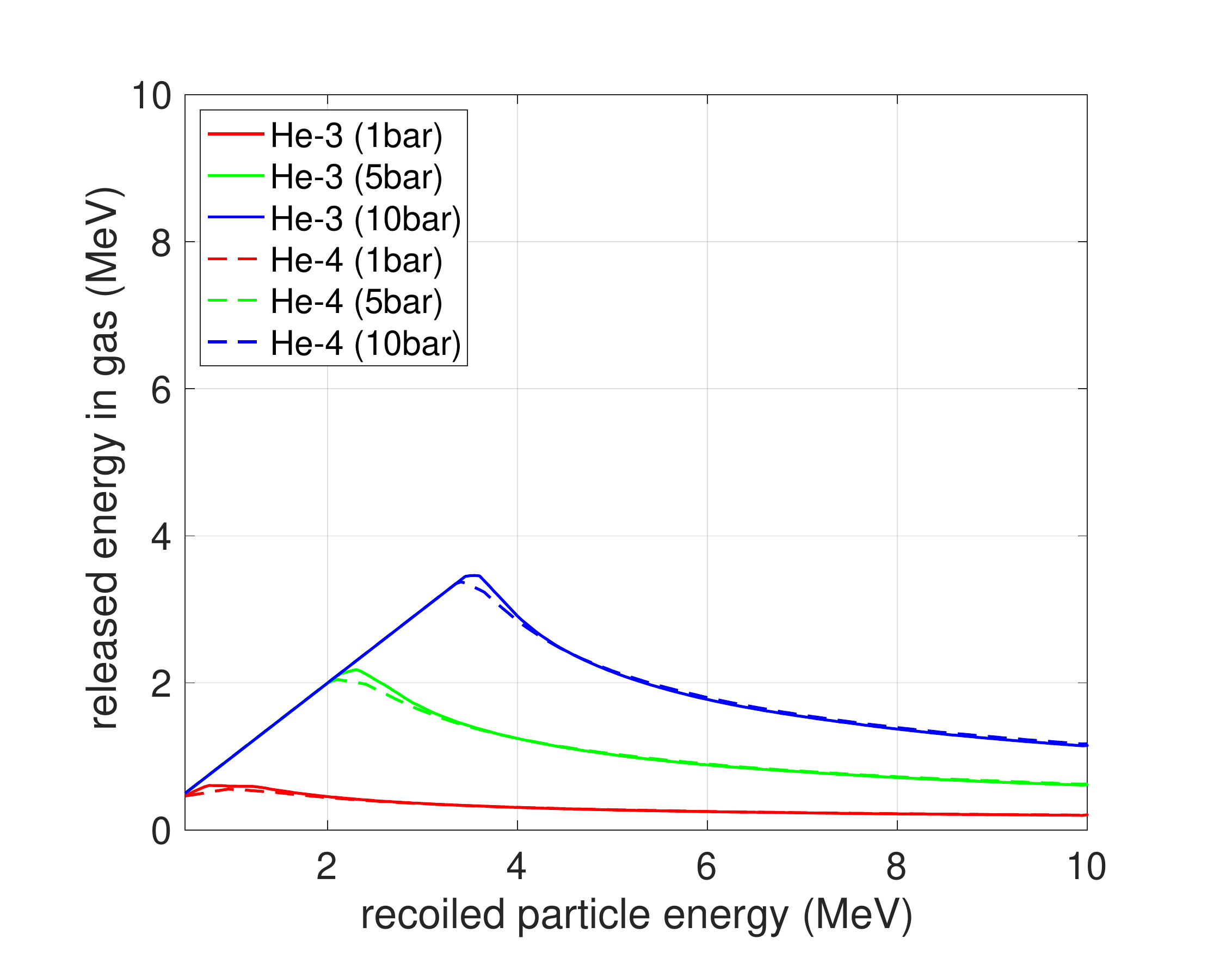}}&
\resizebox{0.31\textwidth}{!}{\includegraphics{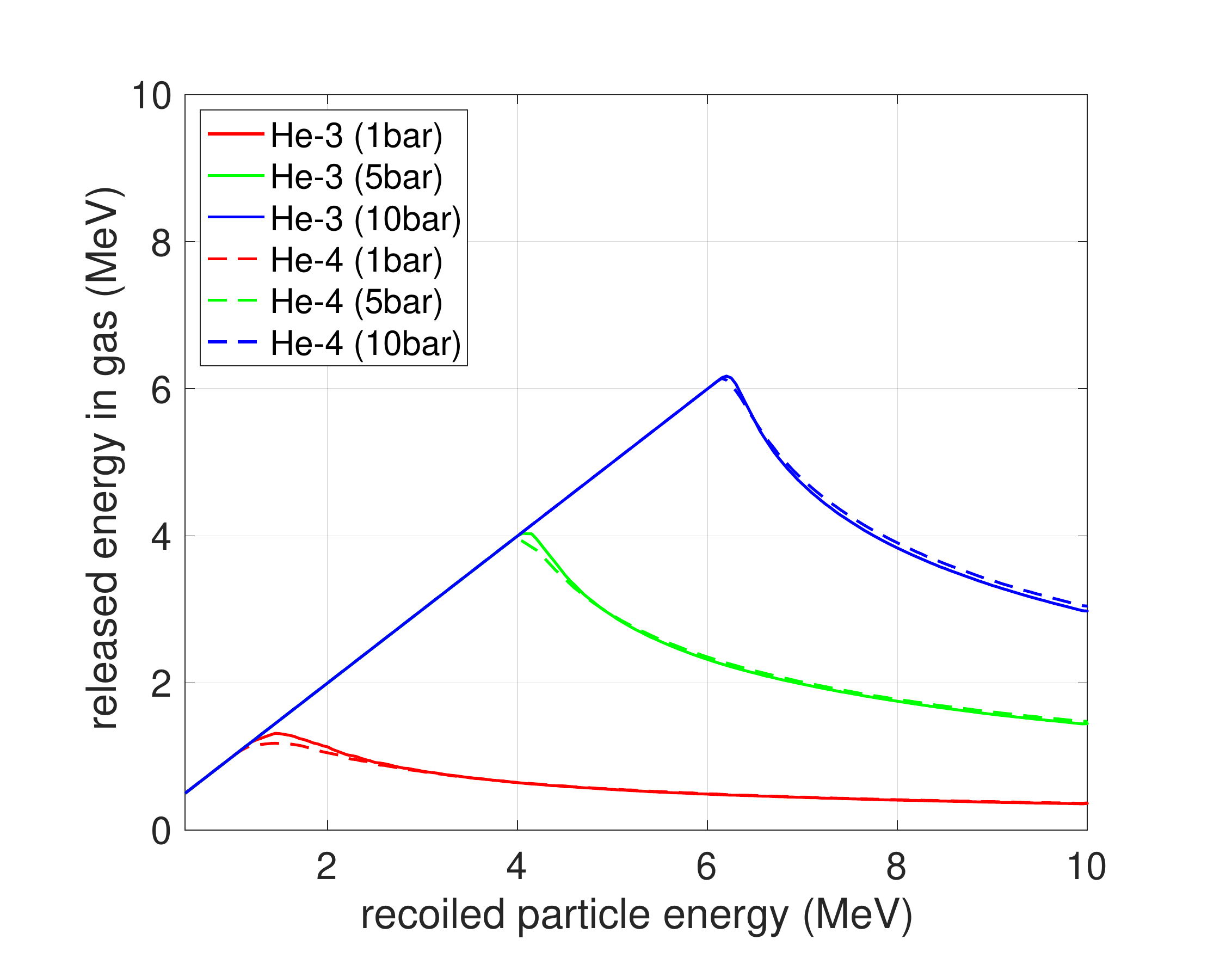}}&
\resizebox{0.31\textwidth}{!}{\includegraphics{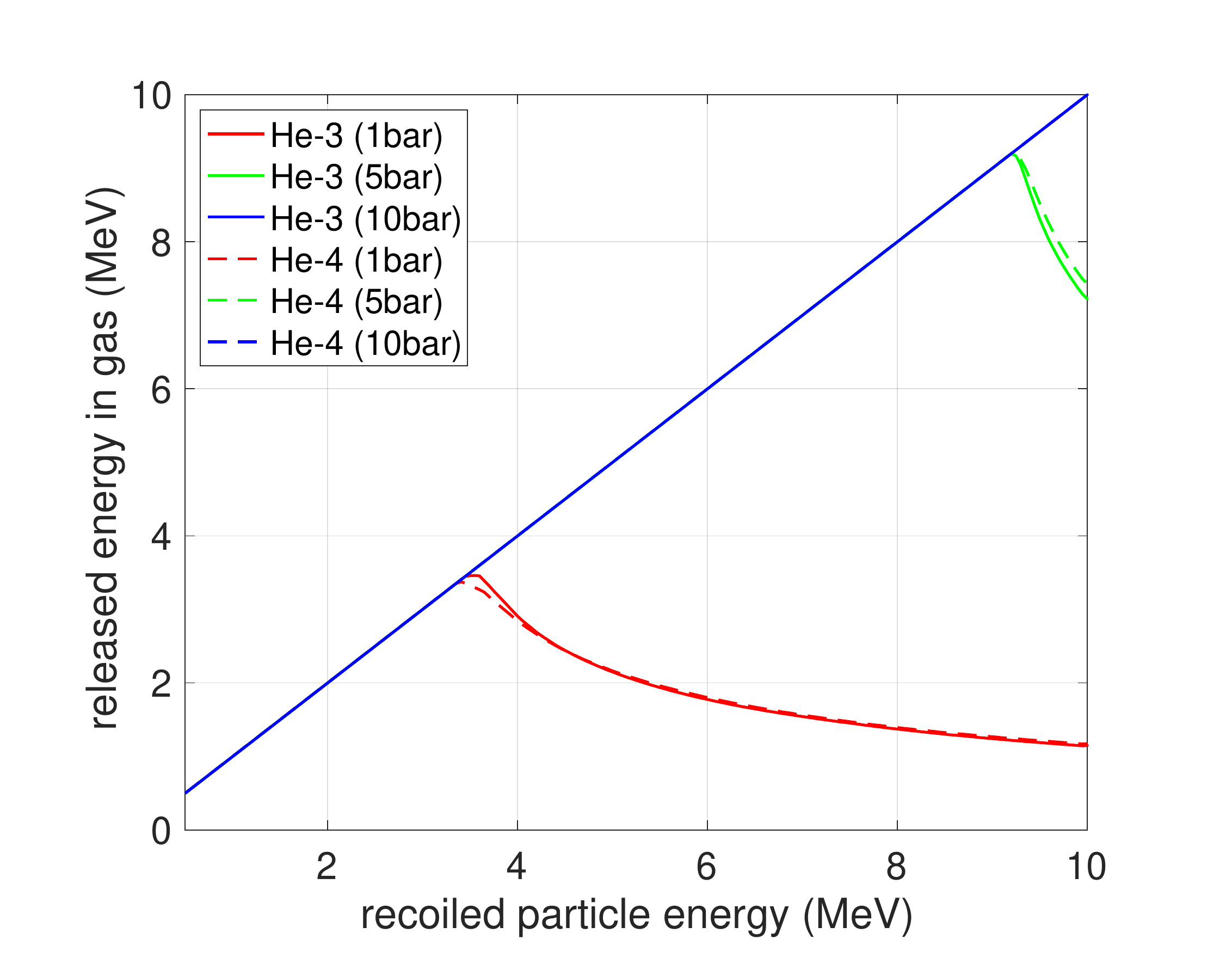}}\\
(a) & (b) & (c)
\end{tabular}
\caption{\footnotesize Released energy of ions ($\mathrm{^3He^{++}}$ in $\mathrm{^3He}$ and $\mathrm{^4He^{++}}$ in $\mathrm{^4He}$) as a function of the energy gained after recoiling for 1, 5 and 10\,bar and a) 1~cm gas depth, b) 2.54~cm and c) 10~cm.}
\label{figrelenerg} 
\end{figure}

Although the mass density of the two gases differs, the ranges of the recoiled nuclei are similar. 
\\ For different thicknesses of the gas volume and pressures one can calculate, from the stopping power, the maximum energy that can be released in the gas. The dependency between the energy a nucleus can release and its recoiling energy is linear if the gas volume is large enough to allow the ion to be fully stopped. Figure~\ref{figrelenerg} shows the relationship between the recoiling energy and the released energy for three different thickness of the gas volume (1\,cm, 2.54\,cm and 10\,cm) and three different pressures of the gases (1, 5 and 10\,bar). 

Given that the ranges in the two media are similar, it is clear that for a given volume and pressure, the energy deposition in the gas for the two isotopes follows the same behaviour. 

To conclude, the two isotopes, for a given volume, pressure and detector geometry, behaves similarly: they show a comparable average total cross-section in the range 1-10~MeV, the interaction of a fast neutron with Helium-3 or Helium-4 has a similar probability to happen, the energy released for a recoiled ion, with given energy, is almost identical; i.e. the range of an $\mathrm{^4He^{++}}$ ions in Helium-4 is similar to the range of an $\mathrm{^3He^{++}}$ ion in Helium-3. The only difference between the two isotopes is the maximum energy that the recoiled nucleus can receive from an impact ($E_{R\,max}$). 
Hence, the order of magnitude for the sensitivity of Helium-3 to fast neutrons is the sensitivity of Helium-4 by just accounting for the different maximum recoil energy (equation~\ref{eq1}), i.e. the energy axis of the Helium-4 PHS has to be shifted by the quantity $\frac{0.75}{0.64}=1.17$. This shift is needed if a non-mono-energetic source of fast neutrons is used, with a mono-energetic beam instead one can use a different imping energy, 1.17 times higher for Helium-4 than Helium-3 to have the same average recoil energy. 
\\ For a given incoming distribution of neutron energies, the PHSs of Helium-3 and Helium-4 (even if shifted), will not only differ for energies below $Q=0.764\,\mathrm{MeV}$, where thermal and epi-thermal neutrons contribute to the spectrum, but also above such energy, where in both detectors only fast neutrons contribute. Assuming an identical total macroscopic cross-section for both isotopes, the Helium-4 cross-section is entirely given by the elastic process whereas, that of Helium-3 is given, on average, by 1/5 by the absorption and 4/5 by the elastic cross-section. The absorption process is responsible of generating events in the PHS at the energies $E_i+Q$ with all possible $E_i$. This will raise the PHS of Helium-3 at energies above $Q=0.764\,\mathrm{MeV}$ by approximately 1/5. This effect will be discussed in section~\ref{resultsec}.
\\Nevertheless, the overall sensitivity of Helium-4, is still a good indication of the sensitivity of Helium-3 because in order to calculate the sensitivity, only the PHS integral (above a given energy threshold) matters. 

\section{Measurements}
\subsection{Experimental setup}
\label{expsetup}
\begin{figure}[htbp]
\centering
\resizebox{0.7\textwidth}{!}{\includegraphics{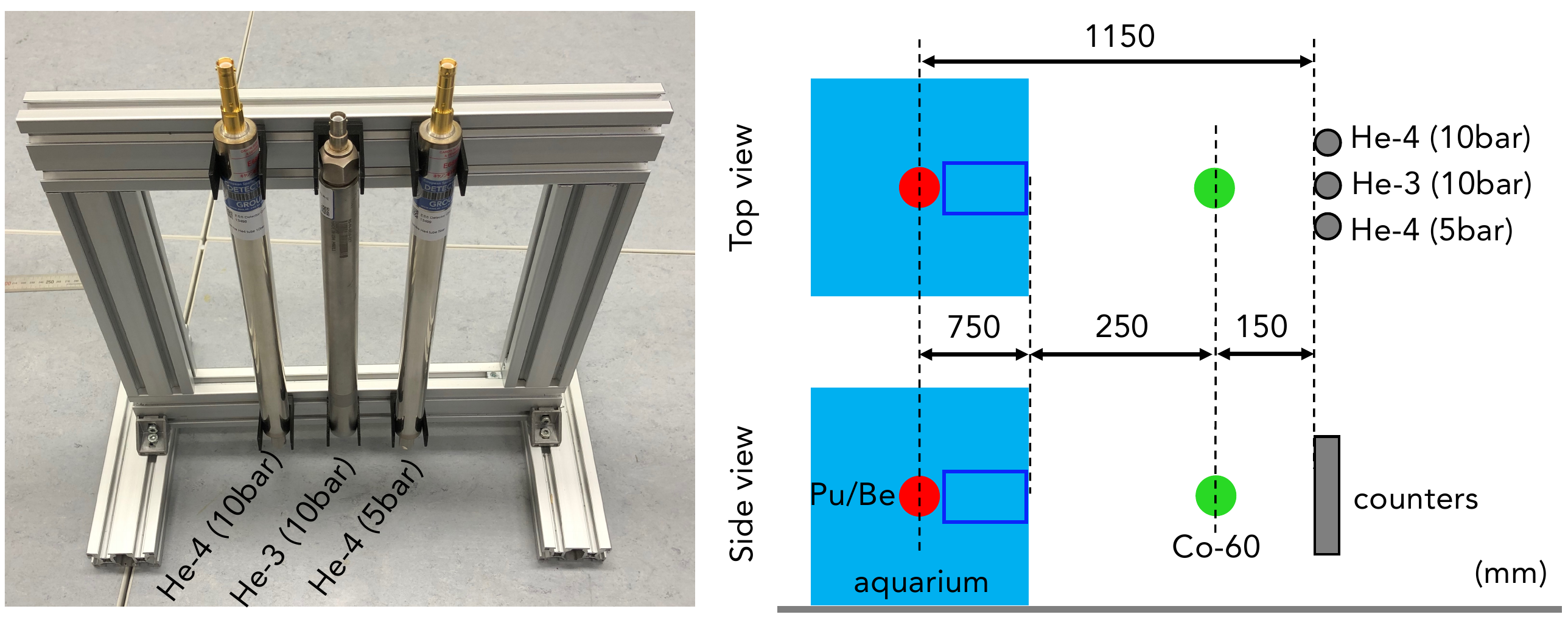}}
\caption{\footnotesize Photo of the stand supporting the three counters used in the measurements (left). Sketch of the experimental set-up (right). Dimensions are in mm.}
\label{figexpsetup}
\end{figure}

Three proportional counters have been used: two Helium-4 tubes, by Toshiba/Canon Electron Tubes \& Devices Co. LTD~\cite{Toshiba}, with partial pressure of 10 and 5 bar respectively and a Helium-3 tube, by General Electric (GE)~\cite{RS}, with partial pressure of 10 bar (see figure~\ref{figexpsetup}). In addition to Helium, in each of the three tubes there is also a small amount ($\approx0.15$~bar) of $\mathrm{CO_2}$. All tubes have an active length of approximately 25~cm and a diameter of 2.54~cm. The tubes are all made of stainless steel which is 0.05~cm thick. 
\\ The efficiency to thermal neutrons of the Helium-3 tube can be calculated according to its pressure and it has also been measured with a mono-energetic neutron beam of 2.5\AA, resulting into a detection efficiency of $\approx96\%$. This value agrees with the calculation within 2\% uncertainty. 
\\ The tubes have been placed side by side and exposed to the same flux (see figure~\ref{figexpsetup}); the measurements have been carried out simultaneously with the three tubes. Each tube is connected to a CREMAT~\cite{EL_cremat} charge amplifier and shaping amplifier in sequence, with overall gain of 14\,V/pC and 1~$\mu$s shaping time. The analogue output of each amplifier is connected to a single channel of a CAEN V1740D digitizer (12 bit, 62.5MS/s)~\cite{EL_CAEN}. The digitizer is equipped with a DPP-QDC (Digital Pulse Processing) firmware; only a value (QDC) proportional to the energy released in each counter is recorded and not the full signal trace. Note that the signals are shaped in time, hence any value among amplitude, pulse integral (QDC) or time-over-threshold (ToT) gives the same information: a value proportional to the energy released in the counter that can be used to build the Pulse-Height-Spectrum (PHS).

The counters have been firstly characterised with cosmic neutrons. These measurement have been carried out at the ESS laboratory Utg\r ard in Lund (SE). 
The measurements with the sources have been performed at the Source Testing Facility (STF)~\cite{SF2,SF1} at the Lund University in Lund (SE). This user facility is equipped with a complete range of neutron and gamma-ray sources for the characterization of detectors. In particular the measurements presented in this manuscript have been performed by using the so-called {\it Aquarium} available at the STF.  
\\ The Aquarium is a custom-designed shielding apparatus for neutron sources, delivering neutron beams at each cavity of this structure. The Aquarium consists of a cube of Plexiglas ($\approx$1.4~m side), filled with about 2650~litres of high-purity water. A neutron source can be placed in its centre. Four horizontal cylindrical apertures of approximately 17~cm diameter are perpendicular to each of the four vertical faces of the cube, provide four uniform combined beams of gamma-rays and neutrons from the source. A sketch of the set-up is shown in figure~\ref{figexpsetup}.
\\ The use of the Aquarium, instead of a bare source, has the effect of decreasing the neutron background detected by the counters. The neutrons that are not exiting the Aquarium through one of its port are absorbed by the large amount of water that surrounds the source. Moreover, the tubes are covered with a 2~mm-thick Mirrobor sheet (only the side of each tube facing the beam port is left uncovered) to further decrease the background neutrons that are scattered by the surrounding environment back into the counters. 

Two sources have been used in the measurements: a $\mathrm{^{238}Pu/^{9}Be}$ (Pu/Be) source of certified activity and neutron yield~\cite{PuBecert} and a Co-60 gamma-ray source of $9\times10^6$~Bq. 
\\The gamma-ray source was used in order to have a comparison with the previous work on gamma-ray sensitivity of Boron-10 and Helium-3 detectors~\cite{MIO_MB2017,MG_gamma,MIO_fastnhe3giac,MIO_fastn}. Co-60 mainly emits gamma-rays at 1.173~MeV and 1.332~MeV with unitary intensity; any other emitted gamma-ray can be neglected for our purposes because of their weak intensity. 
\\ The Pu/Be source has been previously characterised~\cite{PuBe_AmBe} and its neutron emission spectrum, nearly isotropic, ranges from approximately 1 to 10~MeV, with a most probable energy of 3~MeV~\cite{PuBe_AmBe,PuBespectrum}.

\subsection{Results and discussion}\label{resultsec}
\subsubsection{Cosmic neutron fluxes at ground level at ESS}
\begin{figure}[htbp]
\centering
\begin{tabular}{cc}
\resizebox{0.47\textwidth}{!}{\includegraphics{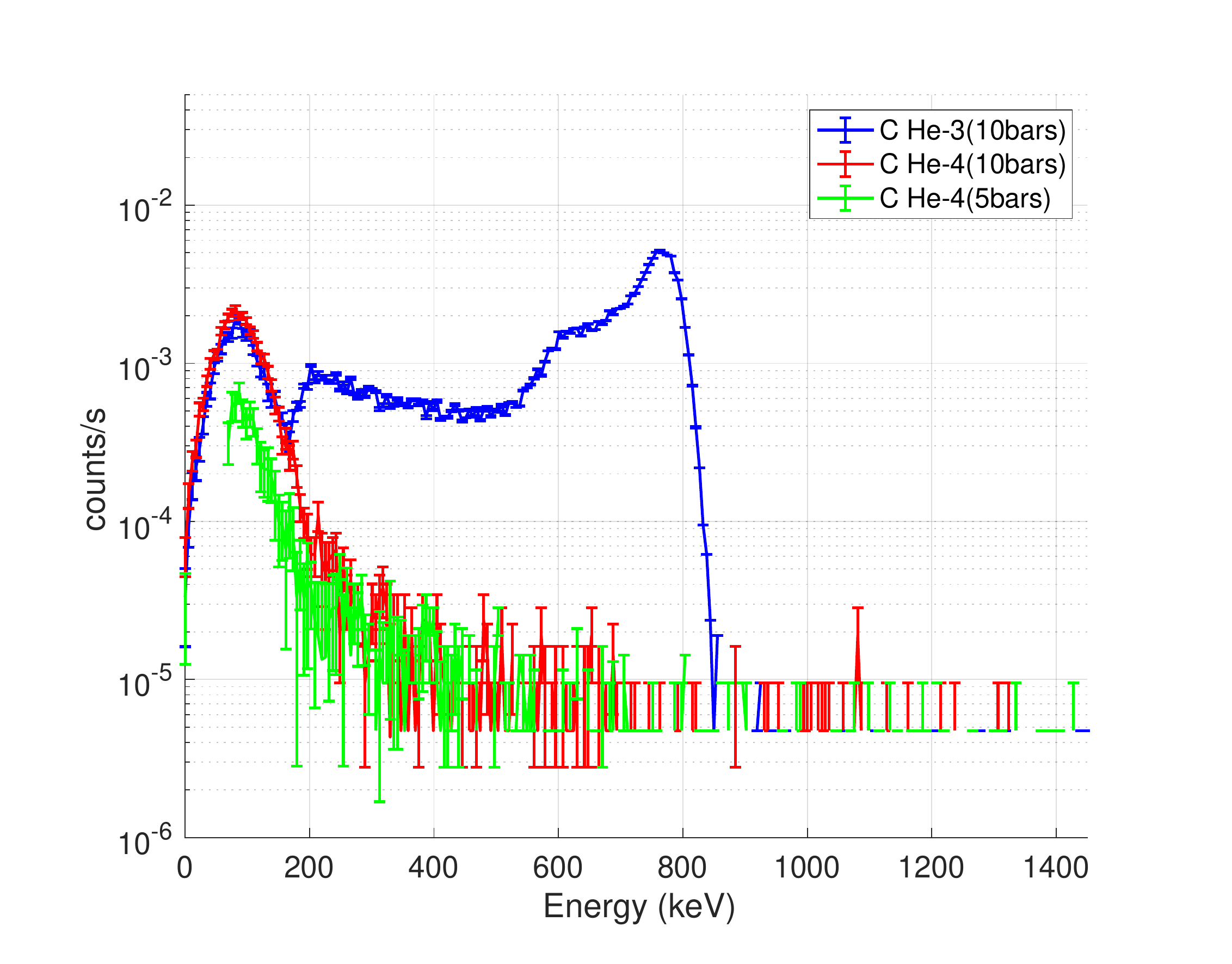}}&
\resizebox{0.47\textwidth}{!}{\includegraphics{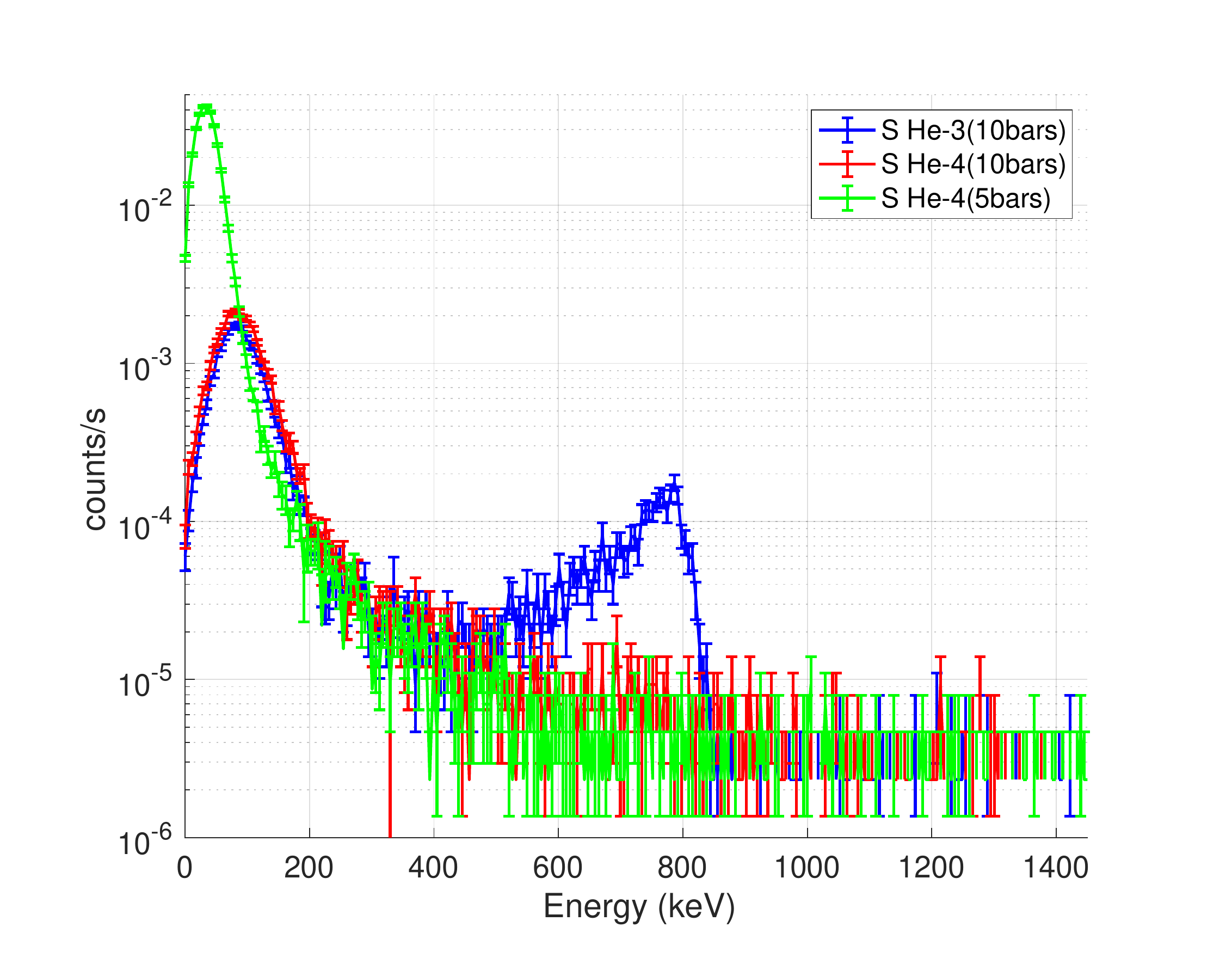}}\\
(a) & (b) 
\end{tabular}
\caption{\footnotesize a) PHS for the three counters in the configuration (C) with the tubes fully exposed. b) PHS for the three counters in the configuration (S) with the tubes covered with a 2-mm thick Mirrobor sheet. The PHS are normalised by the time of the measurements.}
\label{figcosmics} 
\end{figure}

Neutrons are created through the process of cosmic ray spallation when high energy particles collide with atmospheric nuclei. They penetrate much further into atmosphere than the electromagnetic radiations as they are not geomagnetically trapped~\cite{cosmics1}. Cosmic neutrons slow down to thermal energies (appropriately 25~meV) in the atmosphere and neutrons with a very wide energy distribution, spanning from thermal to fast energies, reach the ground level. Those neutrons cannot be distinguished from thermal and fast neutrons generated from the neutron facility. In this section an estimate of the flux of cosmic neutrons at the ground level at ESS is given. These figures can be used in the future to quantify any background generated from cosmic neutrons at ESS. 
\\ Two measurements have been performed with the three tubes recorded simultaneously, one with the bare tubes (C) and one with the tubes fully covered with a 2~mm Mirrobor sheets as a thermal and cold neutron absorber (S). Figure~\ref{figcosmics} shows the PHS, normalised by time, for the three counters in the two configurations. 
\\ The typical spectrum for a Helium-3 counter apart from the peak at $Q=0.764\,\mathrm{MeV}$, has a step structure due to the wall-effect~\cite{DET_knoll}, where the proton (570~keV) and triton (190~keV) energies can be identified. A sharp drop in the spectrum is expected and it extends down to low energies, (a further peak due to the gamma-rays occurs at very low energies.)
The measurement in the configuration (S) serves as a verification of the presence of thermal neutrons in the cosmic incoming flux since the Mirrobor mainly absorbs thermal neutrons and leaves the fast neutron flux unaltered. 

The horizontal axis of the PHS has been converted into energy (keV) from arbitrary channels (QDCs) by using the centre of the peak of the Helium-3 absorption process corresponding to an energy of $Q=0.764\,\mathrm{MeV}$. The hardware threshold in the CAEN system has been set in order to reject the electronic noise, and it corresponds to approximately 80~keV. Each PHS has no sharp cut a low energies but it extends up to zero. This is an effect due to the DPP-QDC firmware used. When the amplitude of the pulses is close to the hardware threshold used to reject noise, the area of such pulses, calculated by the firmware, is affected by a very high uncertainty. If, instead, the amplitudes of the pulses were recorded, no amplitudes below the set threshold would appear in the PHS. The characteristic Helium-3 PHS is highly attenuated in the configuration (S), meaning that the neutrons reaching the Helium-3 counter are efficiently stopped by the Mirrobor sheet. Due to the imperfection of the electronics chain, the Helium-4 tube at 5~bar shows a higher noise with respect to the others.
\\ Table~\ref{tabcosm1} shows the total number of counts for each detector by summing the PHS from a given threshold up to the maximum energy in the spectrum. The PHS integral is shown for 80~keV and for 150~keV which is the threshold that is generally chosen for a Helium-3 counter in normal operation to reject gamma-rays. The rate recorded when no high voltage (HV) is applied to the counters is also shown in order to evaluate the amount of electronics noise above the set threshold. 

\begin{table}[htbp]
\centering
\caption{\footnotesize Rates recorded by the three counters in three configurations and two values for the energy threshold. The rates are shown for both the configurations (C) and (S) and when no high voltage (HV) is applied to the counters in order to evaluate the electronics noise. The relative uncertainty is approximately 40\% for the data without HV and it never exceeds 3\% for the configurations (C) and (S).}
\label{tabcosm1}
\begin{tabular}{clcccr}
\hline\noalign{\smallskip}
        threshold (keV) &  detector & no HV & no shield (C) & shield (S) &  \\ 
        \noalign{\smallskip}\hline\noalign{\smallskip}
  80  & He-3 (10~bar) & $1.59\cdot 10^{-4}$  & 0.174 & 0.032  & Hz \\
        &He-4 (10~bar) &  $1.49 \cdot 10^{-4}$ & 0.035 & 0.037 & Hz\\
        & He-4 (5~bar)   &  0.121 & 0.111 & 0.470 & Hz\\
         \hline
150  & He-3 (10~bar) & $6.94\cdot 10^{-5}$  & 0.148 & $7.13\cdot 10^{-3}$  & Hz \\
        & He-4 (10~bar) & $6.94\cdot 10^{-5}$ & $4.29\cdot 10^{-3}$ & $5.28\cdot 10^{-3}$ & Hz\\
        & He-4 (5~bar)   &  $1.89\cdot 10^{-3}$ &$4.02\cdot 10^{-3}$ & 0.015 & Hz\\
       \noalign{\smallskip}\hline
\end{tabular}
\end{table}

Cosmic neutron rates have been measured elsewhere by the Physikalisch-Technische Bundesanstalt (PTB) through the use of an extended-range Bonner sphere spectrometer~\cite{cosmics4}. The neutron fluence rates (integrated over the total energy range from $10^-9$ to $10^3$~MeV) are given for three different elevations, and at 85~m over the sea level (similarly to the elevation of Lund, Sweden), the rate is about 135~Hz/m$^2$~\cite{cosmics2,cosmics2_b}. In order to compare with the results shown here, the rates table~\ref{tabcosm1} have to be normalised by the area or volume of the detectors. Table~\ref{tabcosm2} shows the rates for the two 10~bar counters with 150~keV threshold applied recorded in configuration (C) and (S) and normalised by the side area of the tube ($6.3\cdot 10^{-3}$~m$^2$) and by their volume ($5.07\cdot 10^{-4}$~m$^3$). \\ For the normalisation by area is used the actual cross-section of the counters that is generally used in neutron experiments which have detector arrays comprised of many such tubes. 

\begin{table}[htbp]
\centering
\caption{\footnotesize Rates for the two 10~bar counters with 150~keV threshold applied recorded in configuration (C) and (S) and normalised by the side area of the tube ($6.3\cdot 10^{-3}$~m$^2$) and by their volume ($5.07\cdot 10^{-4}$~m$^3$).}
\label{tabcosm2}
\begin{tabular}{l|cc|cc|cc}
\hline\noalign{\smallskip}
detector  &  \multicolumn{2} {c} { rate per tube (Hz)} & \multicolumn{2} {c} { rate per area (Hz/m$^2$)} & \multicolumn{2} {c} {rate per volume (Hz/m$^3$)} \\
\cline{2-7} 
               &    no shield (C) & shield (S)      & no shield (C) & shield (S) & no shield (C) & shield (S) \\
\noalign{\smallskip}\hline\noalign{\smallskip}
He-3 (10~bar) & 0.148  & $7.13\cdot 10^{-3}$                        & 23.3       &   1.13     & 292.1   &   14.1    \\
He-4 (10~bar) & $4.29\cdot 10^{-3}$  & $5.28\cdot 10^{-3}$ &  0.676   &     0.84   & 8.47     &    10.4  \\
\noalign{\smallskip}\hline
\end{tabular}
\end{table}

Given that the tubes are filled with 10~bar, a thermal cosmic neutron flux of about 0.029~Hz/(bar$\cdot$ litre) of Helium-3 is expected without shielding and of about 0.0014~Hz/(bar$\cdot$ litre) with shielding. These rates are in agreement with other measured cosmic rates at the FRM-II reactor~\cite{cosmics3}. Moreover, they are also compatible with the measurements carried out by PTB~\cite{cosmics2,cosmics2_b} of 135~Hz/m$^2$. One has to consider that the Helium-3 counter is mainly sensitive to the thermal neutrons as a Bonner sphere spectrometer used without the polyethylene moderator and the Helium-4 detector as a Bonner sphere more sensitive to fast neutrons with a very low efficiency. Hence the rate presented here represents only a fraction of the spectrum measured by PTB. 

In the case of a neutron scattering experiment, the proportional counters are stacked in an array. The figures extrapolated for a single tube, for either a shielded detector or a bare detector, represent an upper limit for the detected cosmic neutron events. In fact, the stacking of counters induces a shielding that each counter has on its neighbours; and this can only reduce the amount of cosmic neutrons reaching a particular counter. 
\\ Moreover, the imperfection of the manufacture of the counters introduces an extra uncertainty, on the gas gain variation, consequently on the energy thresholds, across the tubes. Thus, the cosmic neutron background measured here only represents an order of magnitude of the neutron fluence at ESS to which neutron scattering detectors are exposed. 

\subsubsection{Fast neutron sensitivity of He-3 and He-4 counters}
\begin{figure}[htbp]
\centering
\resizebox{0.9\textwidth}{!}{\includegraphics{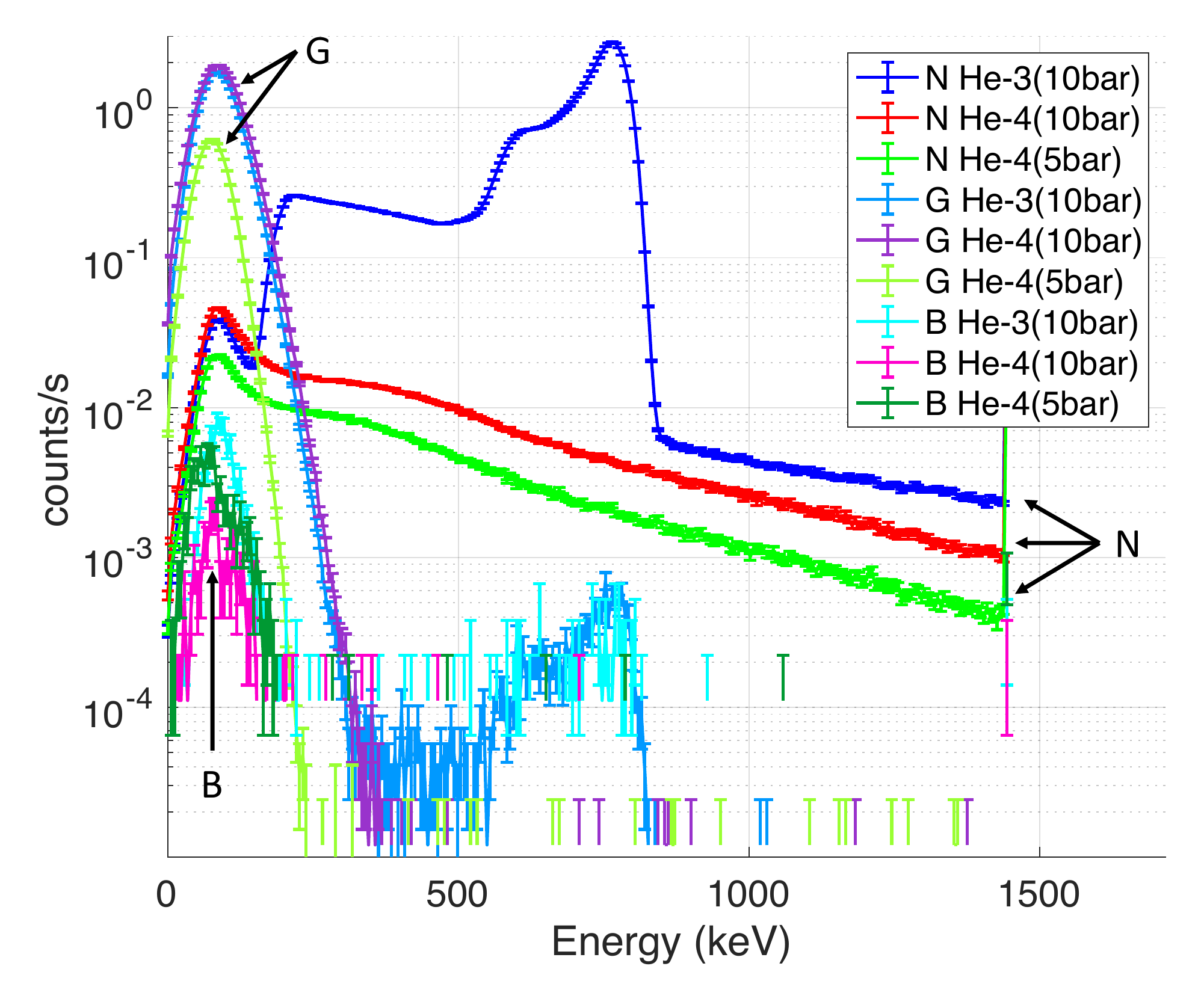}}
\caption{\footnotesize PHS recorded with the three counters in the three different measurements: B (background with no sources), N (Pu/Be) and G (gamma-rays). The PHS are normalised by the relative duration of the measurement.}
\label{figphsall} 
\end{figure}

In this section the absolute sensitivity of a Helium-3 counter is obtained by comparing with a Helium-4 detector. Three measurements have been carried out with the three counters used simultaneously: a background measurement (B), a measurement with the Pu/Be source (fast neutrons - N) and a measurement with a Co-60 source (gamma-rays - G). 
\\ The background (B) has been estimated by removing all radioactive sources from the experimental area. Neutrons are very complicated to shield and for this reason the Pu/Be source has been taken in a separate laboratory area not to influence our background measurement. Figure~\ref{figphsall} shows the PHS recorded in the three configurations (B,N,G) with the three counters. The uncertainty bars in the plot represent the poissonian uncertainty due to the counting statistics. 

The horizontal axis of the PHS has been converted into energy as explained in the previous section. The acquisition system has a input dynamic range limited between zero and two Volt, thus every event with larger amplitude is recorded with the same value indifferently from the real charge deposited; these overflow events are shown in the last bin of the histogram (at an energy of approximately 1450~keV). A gaussian function is used to fit the full energy deposition peak from the neutron capture reaction at Q=0.764~MeV. The Full-Width-Half-Maximum (FWHM) of such a function gives the energy resolution of the counter and it is approximately 60~keV. Thus, any nucleus that recoils with an an energy below this resolution cannot be distinguished from any other neutron which is instead captured. 
\\ By comparing the PHS recorded with the gamma source (G) from figure~\ref{figphsall} one can notice that the Helium-3 tube still shows the characteristic neutron capture spectrum from energies between $\approx 400$ and 800~keV. Although the neutron source was removed from the experimental area, some neutron can still reach the experiential set-up generating these events. These events represent less than $\approx$0.4\% of the total counts in the PHS. 
\\ Table~\ref{tab99} shows the rates recorded with the three tubes in the three different measurements when no further software threshold is applied by software to the PHS. 

\begin{table}[htbp]
\centering
\caption{\footnotesize Rates recorded by the three counters in three measurements: background, fast neutrons (Pu/Be), gamma-rays (Co-60). The recorded rates are all events that trigger the CAEN digitizer with a set hardware threshold which corresponds to approximately 80~keV normalised to the duration of the relative measurement.}
\label{tab99}
\begin{tabular}{lcccr}
\hline\noalign{\smallskip}
         detector & no sources (B) & Pu/Be (N) & Co-60 (G) & \\ 
        \noalign{\smallskip}\hline\noalign{\smallskip}
        He-3 (10~bar) & 0.086 & 70 & 25 & Hz \\
        He-4 (10~bar) & 0.021 & 2.2 & 30 & Hz\\
        He-4 (5~bar)   & 0.064 & 1.1 & 7.5 & Hz\\
       \noalign{\smallskip}\hline
\end{tabular}
\end{table}

\begin{figure}[htbp]
\centering
\begin{tabular}{cc}
\resizebox{0.47\textwidth}{!}{\includegraphics{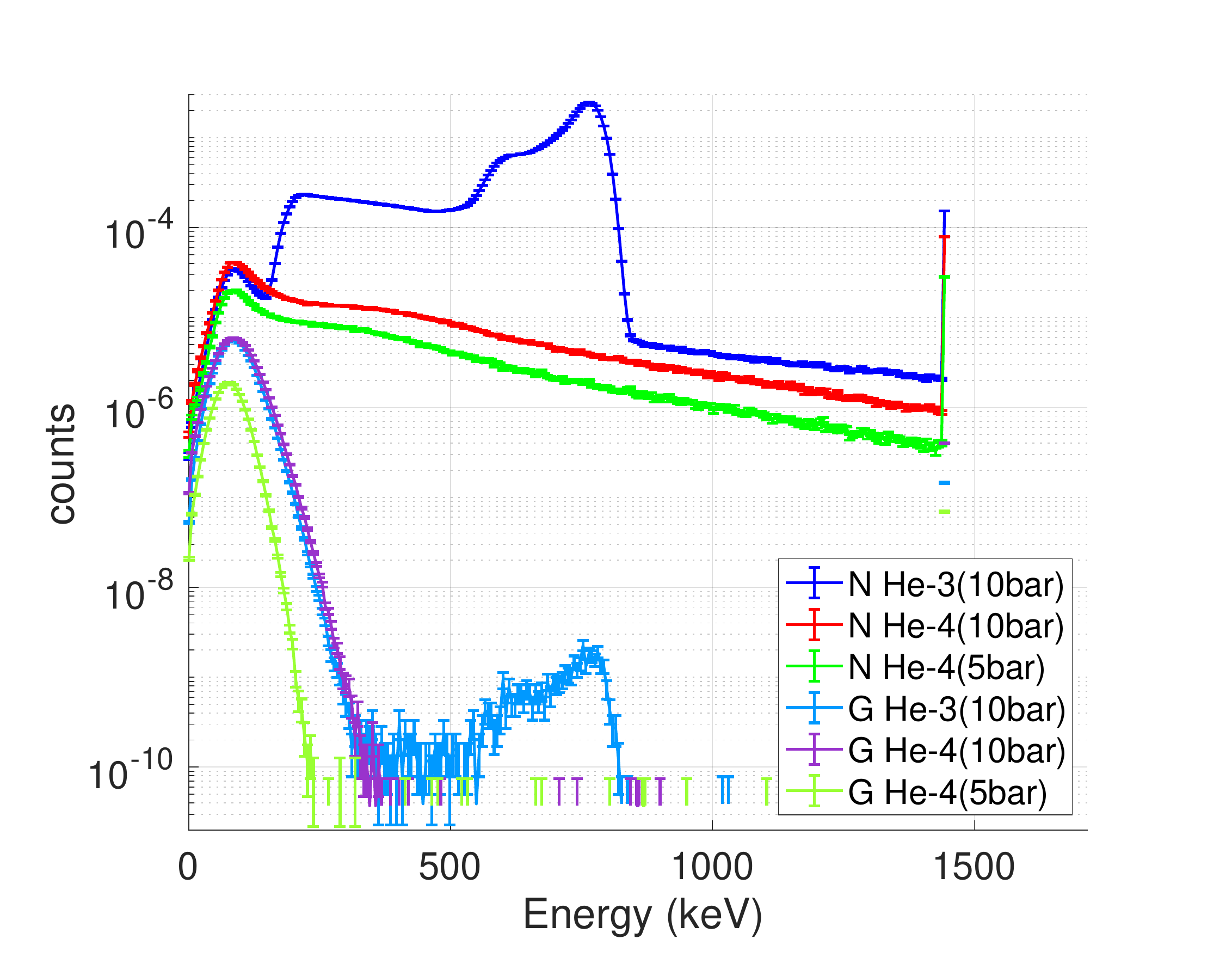}}&
\resizebox{0.47\textwidth}{!}{\includegraphics{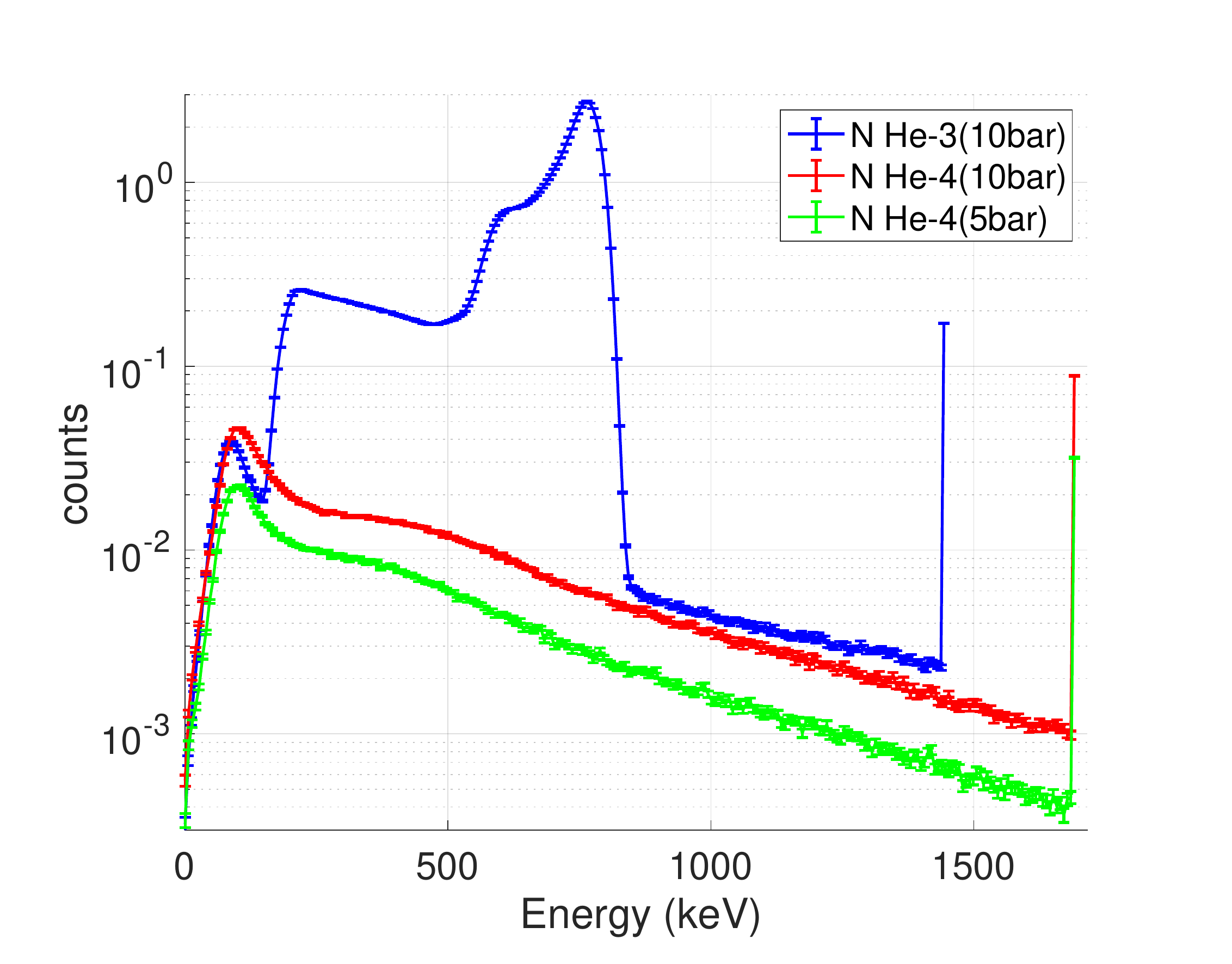}}\\
(a) & (b) 
\end{tabular}
\caption{\footnotesize (a) PHS of the three counters in the configurations N (Pu/Be, neutrons) and G (Co-60, gamma-rays) normalised by the duration of the measurement, solid angle subtended by the source and its activity. (b) PHS of the three counters in the configuration N also normalised and the horizontal axis of the PHS belonging to the Helium-4 detectors also shifted by the ratio $\frac{0.75}{0.64}=1.17$.}
\label{figphsnorm} 
\end{figure}

By comparing the rates in table~\ref{tab99}, the background affects the other measurements by no more than $5~\%$. 
\\ The greatest uncertainty comes from the measurement of the distance between the source and the detectors. We estimate that this can lead to the uncertainty of no more than a factor 2. The encapsulation of the Pu/Be is not the same of that of Co-60. The Co-60 can be considered a point-like source, hence it was possible to place it closer to the tubes than the Pu/Be (subtending a solid angle of 0.23 str). The Pu/Be was instead placed in the Aquarium (subtending a solid angle of 0.005 str) for two reasons: to reduce background from scattered and thermalized neutrons, and because its core, of approximately 1~cm diameter, cannot be considered a single point of emission. By placing it, instead, at more than 1~m from the counters, the uncertainty due the extension of the source is kept below $1\%$ ($\frac{\mathrm{1~cm}}{\mathrm{115~cm}}$).
\\ All figures are given here with their associated poissonian uncertainty due to the counting statistics; in addition to that one has to consider that further uncertainties come from the placement of the sources (a factor 2) and background ($\leq 5~\%$). This does not affect the aim of this work since only an order of magnitude for the fast neutron sensitivity is extracted. 
\\ The sensitivity is the ratio between the detected events, above a given energy threshold, and the total number of impinging particles on the detector. The PHS has to be normalised by the duration of the measurement, the solid angle subtended by the source and by the number of emitted particles by the source in $4\pi$ and unit time (proportional to the activity of the source). Figure~\ref{figphsnorm} (a) shows the normalised PHS for the two configurations N and G.

\begin{figure}[htbp]
\centering
\resizebox{0.9\textwidth}{!}{\includegraphics{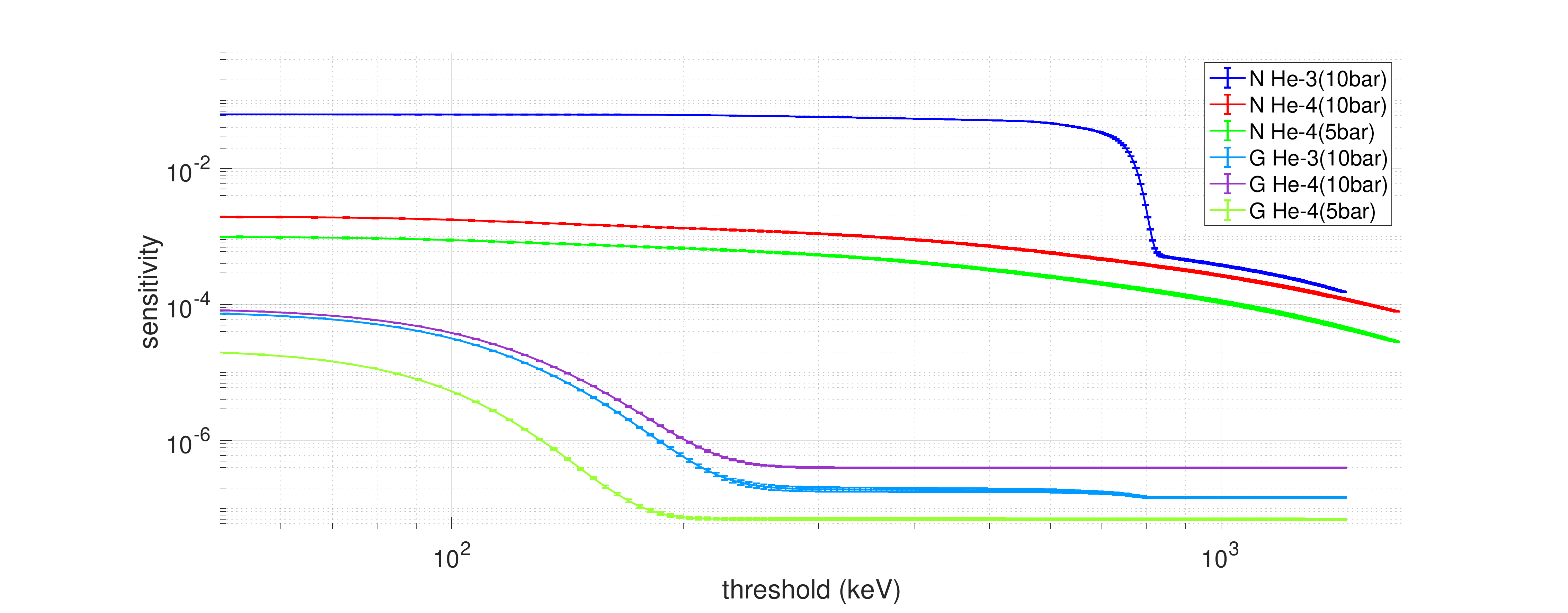}}
\caption{\footnotesize Fast neutron and gamma-ray sensitivity of the three counters as a function of the energy threshold applied to their PHS.}
\label{figsens} 
\end{figure}

As described in section~\ref{theo}, in order to extrapolate the sensitivity of the Helium-3 counter, the horizontal axis of the PHS belonging to the Helium-4 detectors has to be shifted by the ratio $\frac{0.75}{0.64}=1.17$ to account for the different recoiling energy gained by the two isotopes. This is shown in figure~\ref{figphsnorm} (b). Note that, the PHS of the Helium-3 detector, for energies above the full energy deposition peak $Q=0.764\,\mathrm{MeV}$, is approximately 1/5 higher than that of the Helium-4 detector (10~bar); this is in agreement with the considerations discussed in section~\ref{theo}. 
\\ Figure~\ref{figsens} shows the sensitivity as a function of the energy threshold for each tube and for the two configurations N and G. This plot is obtained by summing all counts from a given value of the threshold up to the highest energy in the PHS. The sensitivities are also reported in table~\ref{tabsens} for the values of threshold usually set in normal operation of the counters: 100, 150, 200 and 250~keV. Note that the figures in figure~\ref{figsens} and table~\ref{tabsens} are shown with their associated statistical uncertainties, the uncertainties due to the background (5~\%) and the geometry of the set-up (a factor 2) have to be considered as well when comparing these figures. 

\begin{table}[htbp]
\centering
\caption{\footnotesize Sensitivities for the three counters for the configurations N (Pu/Be) and G (Co-60). Statistical uncertainties are reported as well. In addition to the poissonian uncertainty all figures have systematic uncertainties of a factor 2 due to the precision in the geometrical positioning of the source with respect to the tubes and 5~\% due to the counts due to the environmental background. The sensitivity to fast neutrons (N) of the Helium-3 tube is shown in brackets, it includes thermal and epi-thermal neutron contributions.}
\label{tabsens}
\begin{tabular}{cllll}
\hline\noalign{\smallskip}
        threshold (keV)  & detector & sensitivity N (Pu/Be) & sensitivity G (Co-60)  \\ 
         & & ($\times10^{-3}$) &  ($\times10^{-6}$) \\ 
        \noalign{\smallskip}\hline\noalign{\smallskip}
        100 & He-3 (10~bar) & $(62.15\pm 0.01) $ & $30.13\pm0.03$\\
               & He-4 (10~bar) & $1.721\pm0.002$ & $36.44\pm0.04$\\
               & He-4 (5~bar)   & $0.867\pm0.001$ & $0.488\pm0.01$\\
        \hline
        150 & He-3 (10~bar) & $(61.95\pm 0.01$) & $4.32\pm0.01$\\
               & He-4 (10~bar) & $1.482\pm0.002$ & $6.28\pm0.02$\\
               & He-4 (5~bar)   & $0.751\pm0.001$ & $0.280\pm0.003$\\
        \hline
        200 & He-3 (10~bar) & $(61.10\pm 0.01)$ & $0.508\pm0.004$\\
               & He-4 (10~bar) & $1.315\pm0.002$ & $0.941\pm0.006$\\
               & He-4 (5~bar)   & $0.662\pm0.001$ & $0.073\pm0.001$\\
        \hline
        250 & He-3 (10~bar) & $(59.07\pm 0.01)$ & $0.210\pm0.003$\\
               & He-4 (10~bar) & $1.203\pm0.002$ & $0.433\pm0.004$\\
               & He-4 (5~bar)   & $0.598\pm0.001$ & $0.070\pm0.001$\\
       \noalign{\smallskip}\hline
\end{tabular}
\end{table}

The gamma-ray sensitivity (G) is in agreement with previous works on this matter~\cite{MG_gamma,MIO_MB2017,MIO_fastn}. It can be strongly suppressed by increasing slightly the energy threshold used. The Helium-3 and Helium-4 tubes, at a given pressure (10~bar), have the same gamma-ray sensitivity within the discussed uncertainties. It is very important to properly set the threshold in a neutron experiment because, as for Boron-10 detectors~\cite{MG_gamma}, the gamma-ray sensitivity of a Helium-3 detector scales by orders of magnitude in a very narrow range of the threshold. It is not always possible to identify the ''valley'' cut-off (at the triton (190~keV) escape shoulder) in the Helium-3 PHS and consequently the choice of the threshold becomes more difficult. This is mainly true when a charge division readout method is used for Helium-3 counters, or when higher pressure is used in a Helium-3 tube. 
\\ The sensitivity to fast neutrons (N) of the Helium-3 detector is shown in brackets because it does not reflect the actual tube sensitivity to fast neutrons but it includes the sensitivity to thermal and epi-thermal neutrons as well. 
\\ By comparing the 5~bar and 10~bar tubes, the sensitivity to fast neutrons (for a given threshold) scales linearly with the gas pressure as expected, half pressure of Helium-4 in a counter results in about half sensitivity of the detector. For instance the sensitivity, for 150~keV threshold, is approximately $3\cdot10^{-4}$ per bar $\cdot$ litre. 
\\ The Helium-4 detector at 10~bar indicates the sensitivity of the Helium-3 counter for the same gas pressure and it is of the order of $10^{-3}$. This result is in agreement with both the simulations and experimental results carried out in~\cite{MIO_fastnhe3giac}. 

In a instrumental environment a detector is exposed to a mixed field of radiation, and gamma-rays and fast neutrons fields can easily exceed by orders of magnitude the thermal or cold neutron flux that is relevant for the ongoing experiment. Let us consider the case of an neutron scattering experiment that for instance requires a background rejection below 1~\% of the signal. The absolute gamma-ray flux impinging on the detector must not to exceed $10^{4}$ gamma-rays per thermal neutron detected. And the fast neutron flux must be below $10^{1}$ for an Helium-3-based detector. 
\\ It has been shown that Boron-10-based detectors have a similar gamma-ray sensitivity~\cite{MG_gamma,MIO_MB2017} of Helium-3-based detectors but about 100 times lower fast neutron sensitivity ($10^{-5}$)~\cite{MIO_fastn,MIO_fastnhe3giac}. Thus, for a single thermal neutron detected they can tolerate a fast neutron flux of $10^{3}$ to keep the same background rejection level (below 1~\%).

\section{Conclusions}
New science is enabled by more powerful instruments at neutron sources that are increasing the available neutron flux delivered at the instruments. Higher flux may implicate higher background as well, mostly when heavy shielding is not a viable option due to space limitations. Along with a higher detector performance, such as better spatial resolution and counting rate capability, the signal-to-background (S/B) is a key feature that requires attention to enable the new investigations. Gamma-rays and fast neutrons are the main species of background in a neutron scattering facility, and the flux on a detector of such radiations can easily exceed the thermal (and/or cold) neutron flux which carries the scientific information. Thus, improving the background rejection for a detector is as crucial as increasing the available flux at the instrument. Moreover, cosmic neutrons are an additional source of background that here has been quantified. 

Cosmic neutron energies span a very wide energy range and, when thermalised, they are detected with high efficiency and results into events that are impossible to distinguish from thermal neutron from the neutron instrument. A fluence of approximately 23~Hz/m$^2$ has been found with a bare Helium-3 counter and of about 1~Hz/m$^2$ if covered with a neutron absorber such as Mirrobor. This results into a fluence of about 0.029~Hz/(bar $\cdot$ litre) without shielding and of about 0.0014~Hz/(bar $\cdot$ litre) with shielding. Note that these figures represent an upper limit for the cosmic neutron flux because, in an actual detector when they are arranged in an array, the counters partly screen each other from this radiation. 

Although many detector technologies have been developed to face the new challenges in the past ten years, Helium-3 is still a valid means of detection for some applications. The fast neutron sensitivity of an Helium-3 proportional counter has been verified here through the direct comparison with an Helium-4 detector. In an Helium-3 detector is not possible to directly disentangle the detection of thermal (and cold) neutrons to the detection of fast neutrons without a subtraction method, hence the Helium-4 detector, with identical physical features, was used to measure the sole detection of fast neutrons. In this work the fast neutron sensitivity has been studied between 1 and 10~MeV. 
\\ The fast neutron sensitivity is of the order of $10^{-3}$ (at 10~bar, 2.25~cm diameter) and this results agrees with a previous work which includes simulations and indirect measurements obtained with a subtraction method~\cite{MIO_fastnhe3giac}. As expected the sensitivity to fast neutrons, of approximately $3\cdot10^{-4}$ per bar $\cdot$ litre of Helium-3, scales linearly with the gas pressure. 

The gamma-ray sensitivity has been measured as well and the results agree with the previous works~\cite{MG_gamma,MIO_MB2017,MIO_fastn}; it is of the order of $10^{-6}$ and it scales steeply in a very narrow range of the threshold. 
It is not always possible to identify the ``valley'' cut-off (190~keV) in the Helium-3 PHS, mainly when the counters are readout with a charge division method, or a higher pressure of Helium-3 is used, and consequently the choice of the threshold becomes more difficult and crucial.

It has been shown that Boron-10-based detectors have a comparable gamma-ray sensitivity~\cite{MG_gamma,MIO_MB2017} than that of Helium-3-based detectors, but, on the other hand, about 100 times lower fast neutron sensitivity ($10^{-5}$)~\cite{MIO_fastn,MIO_fastnhe3giac}. 
\\ When a detector is deployed at an instrument, it is exposed to a mixed field of radiation: fast neutron and gamma-ray fluxes can easily exceed the thermal and cold neutron fluxes carrying the scientific information. For a given configuration of an instrument and a targeted S/B, a Boron-10-based and an Helium-3-based detector can tolerate the same gamma-ray background, but a Boron-10 detector can stand 100 times higher fast neutron background than Helium-3-based detectors.  

\begin{acknowledgement}
{ \bf Acknowledgements}
\smallskip
\\ This work was partially supported by the BrightnESS project, Work Package (WP) 4.2 (EU Horizon 2020, INFRADEV-3-2015, 676548) and carried out as a part of the collaboration between the European Spallation Source (Sweden) and the Lund University (Sweden).
\\ The work has been carried out at the Source Testing Facility at the Lund University (Sweden).
\\ The authors would like to thank Toshiba/Canon Electron Tubes \& Devices Co. LTD for their collaboration and for providing the Helium-4 tubes used in this investigation.
\end{acknowledgement}

\bibliographystyle{ieeetr}
\bibliography{BIBLIO}

\end{document}